\documentclass[reprint,aps,prx,showpacs,floatfix,notitlepage,superscriptaddress]{revtex4-2}
\usepackage{epsfig,graphics,amssymb,amsmath,subeqnarray,color,bm,bbm}
\usepackage{graphicx}
\usepackage[toc,page]{appendix}
\usepackage{float}
\usepackage[colorlinks=true,allcolors=blue]{hyperref}

\begin{document}
\title{Non-reciprocal multifarious self-organization}
\author{Saeed Osat}
\affiliation{Max Planck Institute for Dynamics and Self-Organization (MPIDS), 37077 G\"ottingen, Germany}
\author{Ramin Golestanian}
\affiliation{Max Planck Institute for Dynamics and Self-Organization (MPIDS), 37077 G\"ottingen, Germany}
\affiliation{Rudolf Peierls Centre for Theoretical Physics, University of Oxford, Oxford OX1 3PU, United Kingdom}
\date{\today}

\begin{abstract}
A hallmark of living systems is the ability to employ a common set of versatile building blocks that can self-organize into a multitude of different structures, in a way that can be controlled with minimal cost. This capability can only be afforded in non-equilibrium conditions, as evident from the energy-consuming nature of the plethora of such dynamical processes. In the last three decades, synthetic self-assembly has experienced a significant boost with the development of tools to design specific interactions at different scales, from nucleic acids and peptides to proteins and colloids. To achieve automated dynamical control of such self-assembled structures and transitions between them, we need to identify the relevant fundamental aspects of non-equilibrium dynamics that can enable such processes. Here, we identify programmable non-reciprocal interactions as a potential paradigm using which such functionalities can be achieved. In particular, we propose a model that enables a system to learn and retrieve predetermined desired structures and transition between them, thus behaving as a shape-shifter. The learning rule is composed of reciprocal interactions that lead to the equilibrium assembly of the structures, and non-reciprocal interactions that give rise to non-equilibrium dynamical transitions between the structures.
\end{abstract}
                           
\maketitle

\section{Introduction}
In biological systems, small building blocks self-assemble into structures by taking advantage of the thermal agitations to find the matching partners in the medium. To achieve such functionality in an artificial system, the challenge will be to design an interaction matrix between the building blocks so that they self-assemble to predefined units~\cite{Glotzer_Science2004, Hormoz_PNAS2011, Whitelam_ARPC2015, Nguyen_PNAS2016, Scio2020}. A remarkable feature in living systems is a notion of versatility of the building blocks, which allows self-assembly processes to make economical use of the same units under different conditions. Proposals for how to design systems with such a capability have been recently put forward, in particular, in the context of the so-called multifarious assembly mixture model~\cite{Murugan_PNAS2015, Sartori_PNAS2020}. The main challenge in designing multifarious self-assembly is to encode the desired structures in terms of the interactions between the components in the pool so that the structure will be memorized and retrieved  when needed \cite{Keim_RMP2019}. This goal is achieved up to a certain capacity by engineering the specific interactions between the components~\cite{Hormoz_PNAS2011, Murugan_PNAS2015,Sartori_PNAS2020}, in analogy with the Hopfield neural network model \cite{Hopfield_PNAS1982,amit1992book}.

\begin{figure}[b]
\begin{center}
\includegraphics[width=\columnwidth]{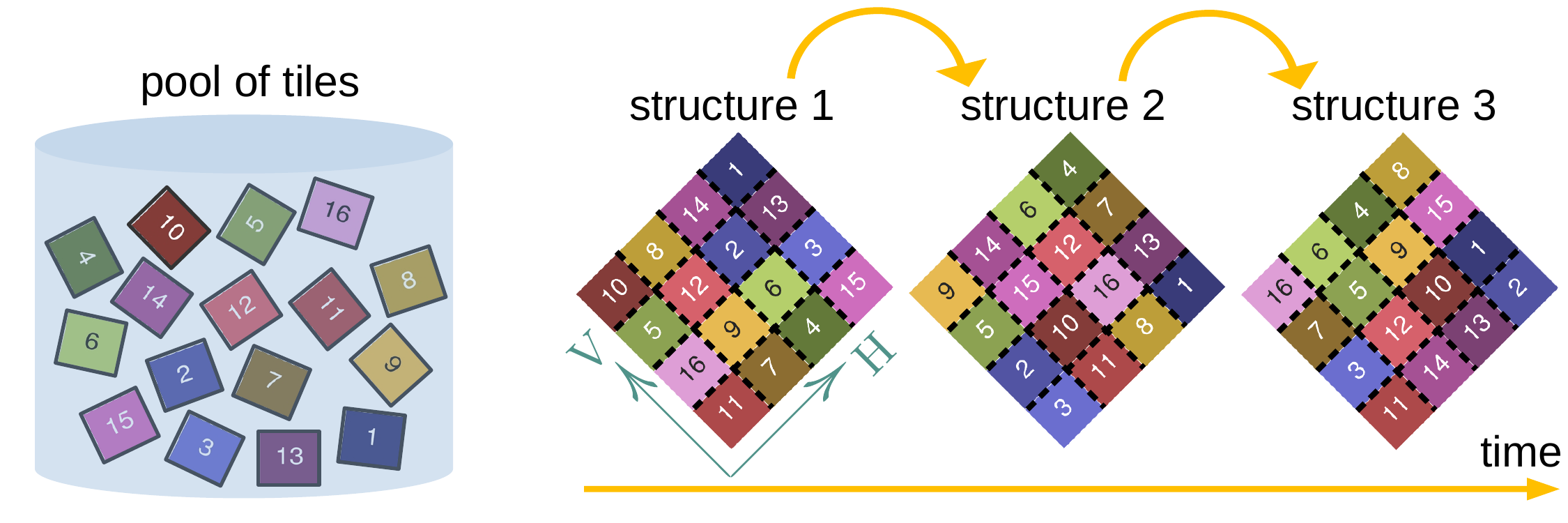}
\caption{Multifarious self-organization with shape shifting. A schematic illustration of the realization of a sequence of programmed transitions between three different predefined target configurations. The 16 unique tiles in the pool can form different desired structures due to the specific reciprocal interactions (dashed-lines). Adding non-reciprocal interactions between the tiles will induce shifting between these structures.}
\label{fig1}
\end{center}
\end{figure}

Living systems have also evolved to be able to choreograph sequential formation of self-assembled structures in time from a common pool of tiles, as schematically shown in Fig. \ref{fig1}. Such self-organization, as exemplified in various stages of the cell cycle \cite{Musacchio2007}, is not possible under equilibrium conditions. Therefore, to design such functionality, we need to identify a relevant non-equilibrium aspect of the process that is capable of driving time-sequenced stochastic dynamics. We propose programmable non-reciprocal interactions as a paradigm that provides sufficient conditions towards achieving this goal. Broken action-reaction symmetry has been recently explored in active matter in the context of non-equilibrium phoretic interactions between catalytically active colloids and enzymes ~\cite{golestanian2019phoretic}, and shown to lead formation of self-propelled active molecules that break time-reversal symmetry \cite{Soto2014}, oscillating active complexes that break time-translation symmetry \cite{Soto2015}, chiral bound-states \cite{Saha2019}, and active phase separation with specified stoichiometry \cite{Jaime_PRL2019, OuazanReboul2021}. Non-reciprocal interactions have been found to lead to rich physical phenomena involving various forms of spontaneous symmetry-breaking in other related non-equilibrium contexts \cite{Uchida2010, Gong_PRX2018, Saha_PRX2020, You2020PNAS,Loos2020, Liu2021, Fruchart2021}, including early work in the context of asymmetric neural networks~\cite{Sompolinsky_PRL1986, Parisi_1986}. We note that a number of strategies have been recently pursued towards experimental realization of shape-shifting soft matter structures \cite{Denkov_Nature2015, Zhang_PRL2016, Haas_PRL2017, Nagarkar2021}.

\begin{figure}[t]
\begin{center}
\includegraphics[width=\columnwidth]{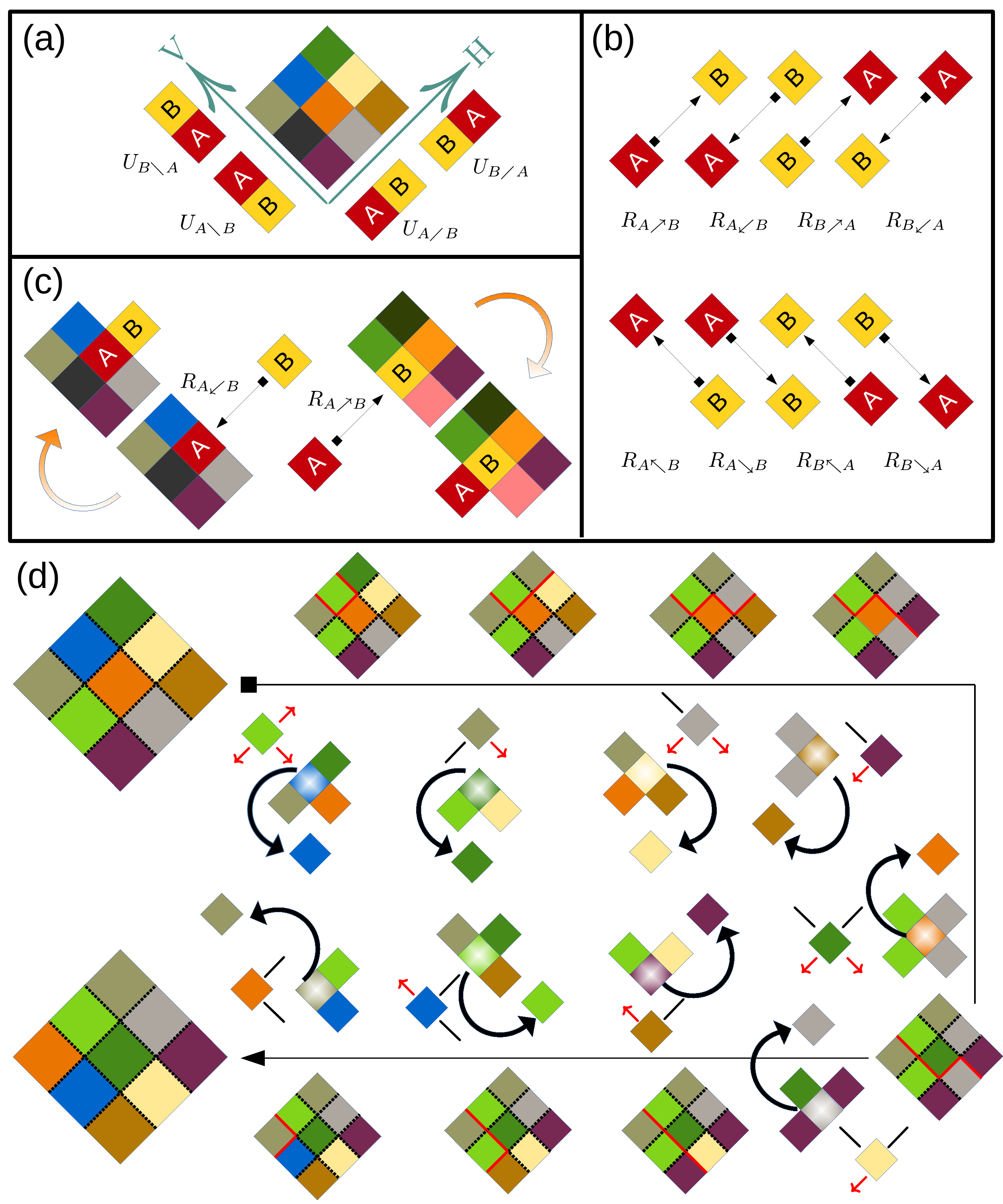}
\caption{Broken action-reaction symmetry and multifarious self-organization model. Panel (a) shows specific reciprocal interactions between tiles A and B on a 2D (H-V) lattice with specific binding energy $U$. Panel (b) illustrates non-reciprocal specific interactions between A and B, and the corresponding interactions $R$. Panel (c) shows non-reciprocity in the presence of a structure; A approaching B is different from B approaching A. Panel (d) gives a schematic representation of shape shifting structures leveraging non-reciprocal interactions. Two desired structures are random permutations of $3 \times 3$ lattice configurations composed of nine distinct tiles. The non-reciprocal interactions induce transitions to the next structures. For arriving tiles, the red arrows and the black lines  show non-reciprocal and reciprocal interactions with the neighbouring tiles, respectively. The red boundary that emerges, grows, and finally disappears, represents the specific bonds that are lost due to the red-arrows. This line shows how a small seed of the new structure emerges due to the non-reciprocal interactions, and eventually conforms to the full structure with the help of reciprocal interactions.}
\label{fig2}
\end{center}
\end{figure}

Here we introduce the non-reciprocal multifarious self-organization model by incorporating non-reciprocal interactions into the equilibrium multifarious self-assembly model. We show that this non-equilibrium model is capable of inducing the shape shifting property in the system in conjunction with the multifarious self-assembly, as depicted in Fig. \ref{fig1}. We aim to find the best parameter space to realize this new shape-shifting regime and to characterize its properties. We show that shape shifters can be observed in a subset of the parameter space where multifarious assembly occurs. We find that the strength of the non-reciprocal interactions provides us with a control parameter to convert any selected multifarious self-assembly regime to multifarious self-organization regime. We characterize the new shape-shifting behaviour by probing the frequency of shifts and the capacity of realizing a cycle composed of a sequence of length~$L$. 

\section{Non-reciprocal multifarious self-organization model} \label{sec:NRMSOmodel}
\begin{figure*}[t]
\begin{center}
\includegraphics[width=0.9\textwidth]{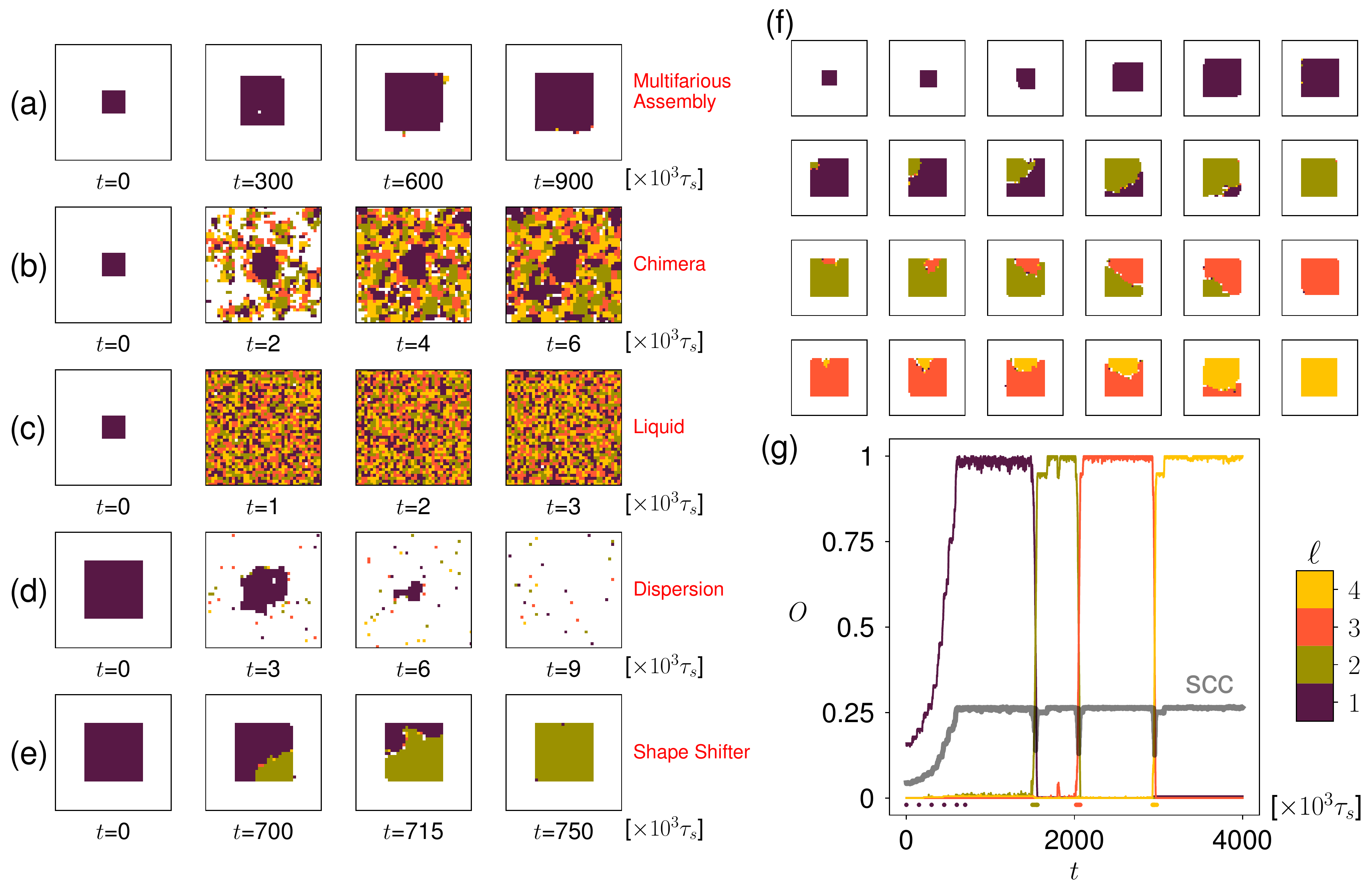}
\caption{Shape-shifting self-organization. A structure is defined as a random arrangement of 400 tiles in a $20\times20$ square. Four desired structures are made by random permutation of the tiles and are encoded through reciprocal [Eq. \eqref{eqn:U}] and non-reciprocal [Eq. \eqref{eqn:R}] matrices. The shifting sequence is selected as \{$S^{(1)} \rightarrow S^{(2)}$, $S^{(2)} \rightarrow S^{(3)}$, $S^{(3)} \rightarrow S^{(4)}$\}. Panels~(a)-(e) show possible outcomes of the simulation. Panel~(f) shows the more detailed transition path of the structures. Non-reciprocal interactions drive the non-equilibrium process that induces the transition between the structures. Starting from the small initial seed of the first structure, the system retrieves the corresponding structure and then shifts to other structures in the sequence. In panel~(g) overlap of the system with each of the patterns is shown as a function time. The dots under the curves show the times at which snapshots have been recorded in panel~(f). The thick grey line shows the size of the largest specifically connected cluster in the system. The $(\text{seed size}, \mu, \varepsilon,\lambda)$ values corresponding to panels (a)-(f) are: (8$\times$8, -19, 11, 7), (8$\times$8, -10, 8, 5), (8$\times$8, 0, 3, 5), (20$\times$20, -10, 4, 0), (20$\times$20, -19, 11, 7), (8$\times$8, -18, 10, 6.5).}
\label{fig3}
\end{center}
\end{figure*}

Equilibrium self-assembly processes enable the spontaneous formation of predetermined structures from a pool of tiles. This can be achieved by designing directional interaction potentials that cover different encounters of two different tiles on a lattice, as shown in Fig. \ref{fig2}(a). These interaction potentials are reciprocal, in the sense that any directional bonding incurs a given energy cost independently of how the tiles come to the binding arrangement. With such interactions, we can achieve self-assembly to a desired structure starting from an initial trigger, be it a small seed, enhanced concentration of some tiles, etc. However, this process will come to a stop once the self-assembled structure is achieved, and remain as such until the next trigger is introduced. Our aim here is to incorporate the ability in the system to autonomously switch from one self-assembled structure to the next. To this end, we define non-reciprocal interactions [see Fig. \ref{fig2}(b)] by attributing different weights for every bond formation depending on which tile is being added and which tile is already a part of the cluster, as shown schematically in Fig. \ref{fig2}(c). Such non-reciprocal interactions give rise to a shape-shifting property amongst the ensemble of possible self-assembled structures from the pool of available tiles, as schematically described in Fig. \ref{fig1}. An example of how such a transition between two different self-assembled structures can occur is shown in Fig. \ref{fig2}(d), where every step is initiated by preferred directional non-reciprocal interaction between an incoming tile and the existing cluster. We show how the interactions can be tuned such that many such transitions occur in a predefined sequence (see Fig. \ref{fig3}).

\begin{figure*}
\begin{center}
\includegraphics[width=0.98\textwidth]{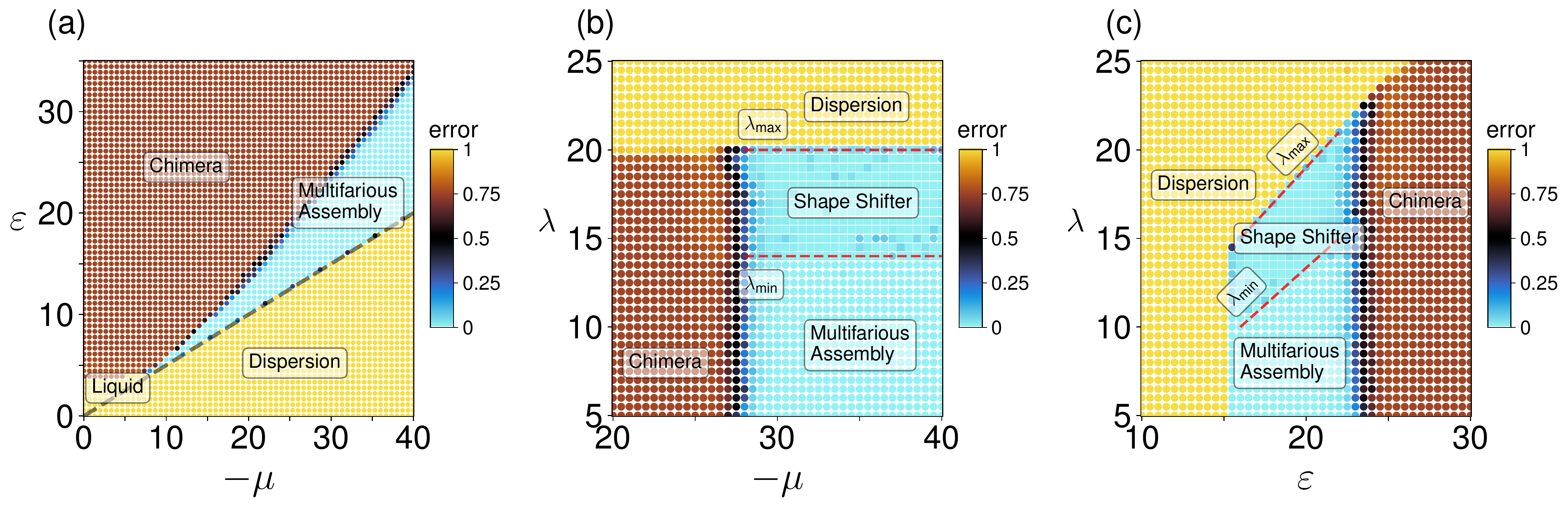}
\caption{The emergence of shape shifter behaviour within the multifarious assembly regime. Four different structures of size $40\times40$ are encoded through reciprocal [Eq. \eqref{eqn:U}] and non-reciprocal [Eq. \eqref{eqn:R}] interactions as a sequence of length three. Each point is an average of error over 10 independent realizations of the system measured after $4\times10^6\tau_s$ time steps. Different forms of self-assembly are obtained starting from the first structure as an initial seed. Panel~(a) shows the self-assembly error for the equilibrium case ($\lambda=0$). Error delineates different modes of self-assembly. For fixed $\varepsilon=21$ and $-\mu\in[20,40]$ we systematically scan the ($\lambda$-$\mu$) parameter space in (b). The square markers correspond to the points that undergo at least one shift. Increasing $\lambda$ beyond the threshold value $\lambda_{\text{min}}$ induces shifting between the structures and converts the multifarious assembly regime to a shape-shifting regime (square markers). Strong non-equilibrium drive for $\lambda>\lambda_{\text{max}}$ makes the initial seed unstable and results in a dilute dispersion. Panel~(c) is the same as panel (b) but for ($\lambda$-$\varepsilon$) parameter space for a fixed~$\mu=-30$ and $\varepsilon\in[10,30]$. Grey dashed line shows the stability limit of the initial structure corresponding to the $2\varepsilon> -\mu$ condition. Red dashed lines are to guide the eye through the limiting values of~$\lambda$.} 
\label{fig4}
\end{center}
\end{figure*}

\subsection{Reciprocal interactions}
We consider $m$ desired structures with each being a random permutation of $M$ components or tiles. For simplicity let us arrange the first $q$ structures in a queue that we label as $S^{(1)}$ to $S^{(q)}$, which defines the shifting sequence. The goal is to define an interaction matrix that enables self-assembly of each of the target structures and the realization of the shifting sequence. We assume that each pair of neighbouring tiles in the desired structures has a specific interaction and define these reciprocal interactions imposed by structure $\ell$ as 
\begin{align}
  U_{A \square B}^{\text{r}} = \begin{cases}
    -\varepsilon, & \text{if } A \square B \in  {\cal I}^{\text{r}}, \\
    0, & \text{otherwise}, 
  \end{cases} 
  \label{eqn:U}
\end{align}
in units of the thermal energy $k_{\rm B} T$, where $\square \in \{ \diagdown , \diagup \}$ represents a specific reciprocal interaction [see Fig. \ref{fig2}(a)]. Moreover, ${\cal I}^{\text{r}} \equiv 
I^{\text{r}} (S^{(1)}) \cup I^{\text{r}} (S^{(2)}) \cup \dots \cup I^{\text{r}} (S^{(m)})$, is the set of all specific interactions between the tiles imposed by $m$ desired structures, with $I^{\text{r}} (S^{(X)})$ being the set of all specific interactions in the structure $\ell$, namely
\begin{align}
    I^{\text{r}} (S^{(\ell)}) = \bigcup_{\substack{ {\left \langle \alpha,\beta \right\rangle } }} \; 
    S^{(\ell)}_{\alpha}  \square S^{(\ell)}_{\beta} ,
\end{align}
where $\alpha$ and $\beta$ are representatives of lattice coordinates $(i,j)$ running over nearest neighbours, and $\square \in \{ \diagdown , \diagup \}$. 
To describe the configuration space of the system, we can define a Potts configuration variable 
$\sigma_{\alpha}=0, 1, 2, \dots ,M$, with $\sigma_{\alpha}=0$ 
representing an empty slot and the others describing the corresponding tile species. Using the interaction potential and the configuration variables, one can define a (classical generalized) Hamiltonian for the system as 
\begin{align}
{\cal H}=\sum_ {\left \langle \alpha,\beta \right\rangle } 
U_{\sigma_{\alpha} \square \sigma_{\beta}}^{\text{r}} -\mu \, n, 
\end{align}
where 
$\mu$ is the chemical potential of the tiles (assumed to be the same for all species), $n=\sum_{\alpha} \left(1-\delta_{0,\sigma_{\alpha}}\right)$ 
representing the total number of tiles in every given configuration, and $\square \in \{ \diagdown , \diagup \}$. As usual, the chemical potential controls the average density of the titles in the system.

The reciprocal interaction $U$ asserts that two components specifically interact if the interaction is favoured at least by one of the structures~\cite{Murugan_PNAS2015, Sartori_PNAS2020}. This simple interaction rule makes multifarious self-assembly model an associative memory capable of retrieving stored structures starting from an initial seed or any similar trigger. With appropriate tuning of the model parameters (the energy scale $\varepsilon$, the number of components $M$, the number of memorized patterns $m$, and the chemical potential $\mu$) one can achieve an equilibrium self-assembly machine reminiscent of the Hopfield neural network~\cite{Murugan_PNAS2015, Bisker_PNAS2018, Zhong_JSP2017}.

\subsection{Non-reciprocal interactions}
Addition of non-reciprocal flavour to specific interactions turns the equilibrium multifarious self-assembly model into the non-equilibrium multifarious self-organization model with new shape shifting property. Inspired by recent diverse physical models with non-reciprocal interactions \cite{Uchida2010, Gong_PRX2018, Saha_PRX2020, You2020PNAS,Loos2020, Liu2021, Fruchart2021,Sompolinsky_PRL1986,Parisi_1986}, we introduce non-reciprocal interactions between the tiles as follows. We define
\begin{align}
  R_{A \blacksquare B}^{\text{nr}}= &\begin{cases}
    \lambda, & \text{if } A \blacksquare B \in  {\cal I}^{\text{nr}}, \\
    0, & \text{otherwise},
  \end{cases}
  \label{eqn:R}
\end{align}
where $\blacksquare \in \{ \searrow, \nwarrow, \nearrow, \swarrow \}$ represents all possible specific non-reciprocal interactions [see Fig. \ref{fig2}(b)]. The set of all such interactions between the tiles needed to realize the shifting sequence \{$S^{(1)} \rightarrow S^{(2)}$, $S^{(2)} \rightarrow S^{(3)}$, \dots, $S^{(q-1)} \rightarrow S^{(q)}$\} is denoted by ${\cal I}^{\text{nr}} \equiv 
I^{\text{nr}} (S^{(1)} \rightarrow S^{(2)}) \cup I^{\text{nr}} (S^{(2)} \rightarrow S^{(3)}) \cup \dots \cup I^{\text{nr}} (S^{(q-1)} \rightarrow S^{(q)})$. Here, $I^{\text{nr}} (S^{(\ell)} \rightarrow S^{(\ell+1)})$ is the set of specific non-reciprocal interactions needed for the realization of the transition $S^{(\ell)} \rightarrow S^{(\ell+1)}$, which is defined as 

\begin{align}
&I^{\text{nr}} (S^{(\ell)} \rightarrow S^{(\ell+1)}) = \bigcup_{\substack{ i,j }} \; 
\Big\{ \;
    S^{(\ell)}_{i-1,j}  \swarrow S^{(\ell+1)}_{i,j},  \\
    &S^{(\ell+1)}_{i,j}  \nearrow S^{(\ell)}_{i+1,j}, \; 
    S^{(\ell+1)}_{i,j}  \searrow S^{(\ell)}_{i,j-1}, \;
    S^{(\ell)}_{i,j+1}  \nwarrow S^{(\ell+1)}_{i,j} \nonumber 
\;\Big\}.
\end{align}
The asymmetric interaction matrix $R^{\text{nr}}$ contains those specific non-reciprocal interactions that are favoured by at least one of the transitions.

\begin{figure}
\begin{center}
\includegraphics[width=1.\columnwidth]{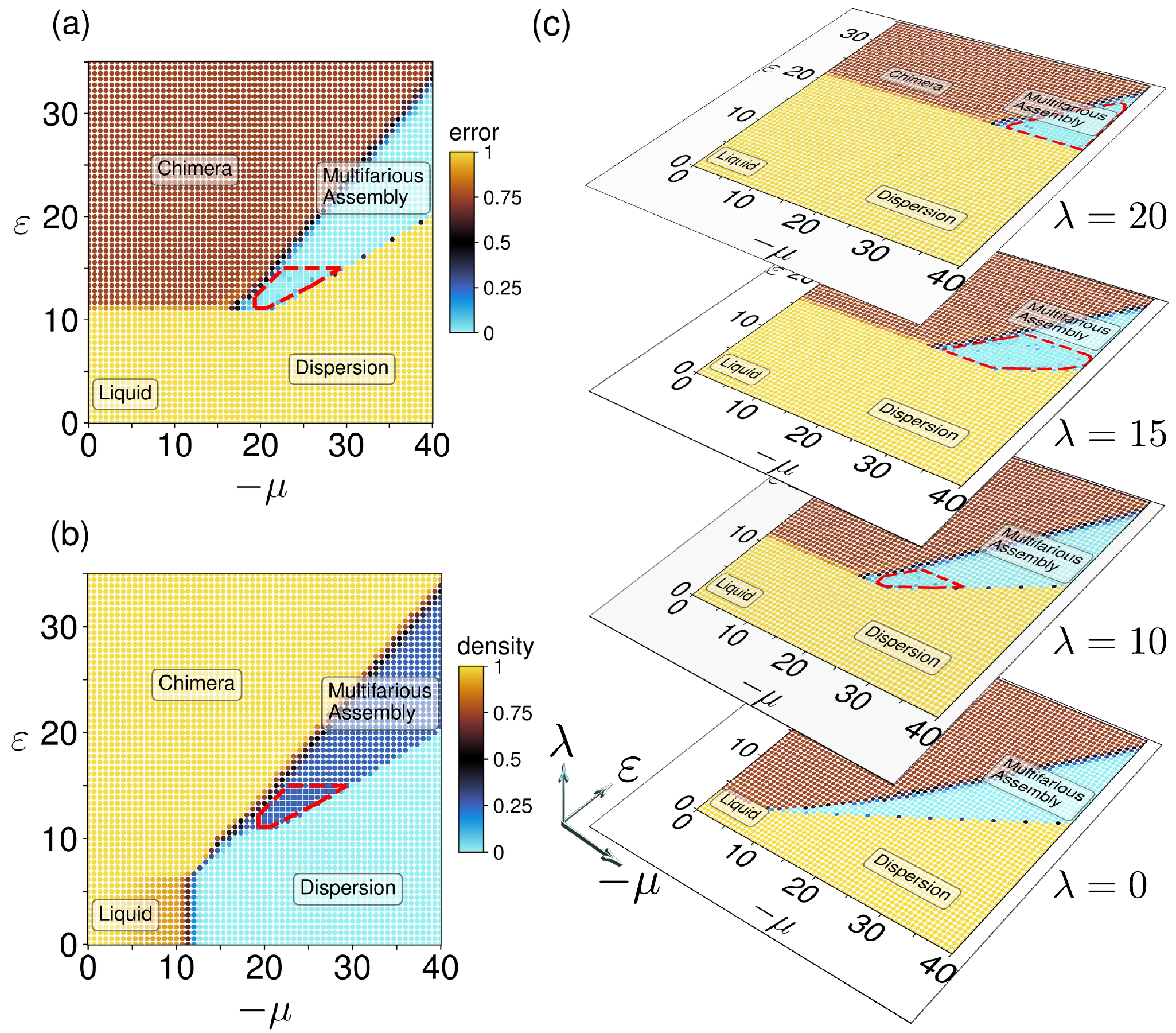}
\caption{($\varepsilon$-$\mu$) parameter space. Four different structures of size $40\times40$ are encoded through reciprocal [Eq. \eqref{eqn:U}] and non-reciprocal [Eq. \eqref{eqn:R}] interactions as a sequence of length three. Panel~(a) shows the average self-assembly error for $\lambda=10$. The red dashed line is the convex hall of the points that correspond to $\text{error}<0.05$ and have undergone at least one shift. Panel~(b) is the same as (a) but for density at the final configuration. Panel~(c) shows the emergence of a shape shifting domain in the multifarious assembly region of the phase diagram. In all the plots, each marker point (corresponding to error or density) is obtained by averaging over 5 independent realizations of the system and after $4\times10^6\tau_s$ time steps.} 
\label{fig5}
\end{center}
\end{figure}

\subsection{Simulation Method} 
The introduction of the non-reciprocal interactions into the multifarious self-assembly model renders the problem inherently non-equilibrium. As such, a faithful treatment of the stochastic dynamics will require the use of an appropriate Master equation formalism. To help highlight the connection with the equilibrium multifarious self-assembly model, however, we have chosen to use a generalized Monte Carlo scheme in which we have incorporated the non-reciprocal interactions in the spirit of kinetic Monte Carlo algorithms. Our specific implementation can be justified with the assumption of separation of time sales between the process of self-assembly and the shape-shifting transitions. 

In the lattice realization of our model, the whole system is defined as a square lattice of size~$2\sqrt{M} \times 2\sqrt{M}$, in which desired structures in the form of 2D square lattices of size $\sqrt{M} \times \sqrt{M}$ would be embedded. We make use of fully-heterogeneous and zero-sparsity condition, i.e. each component should appear only once in each structure~\cite{Sartori_PNAS2020} thus each structure is a random permutation of the tiles in the square lattice. We follow a generalized version of the grand canonical Monte Carlo simulation as would have been implemented for the Hamiltonian ${\cal H}$, with the following generalization in the acceptance rate of every step. At each Monte Carlo step, a random lattice point $(i,j)$ is chosen and its component $\sigma_{i,j}$ is changed to another random component $\sigma^{\prime}$ with probability 
\begin{align}
    p={\rm min}\big\{ 1, \exp\left(\Lambda-\Delta {\cal H}\right)\big\},
\end{align}
where 
\begin{align}
    \Lambda_{i,j}= R_{\sigma_{i-1,j} \swarrow \sigma^{\prime}}^{\text{nr}}+
          R_{\sigma^{\prime} \nearrow \sigma_{i+1,j}}^{\text{nr}}+ 
          R_{\sigma^{\prime} \searrow \sigma_{i,j-1}}^{\text{nr}}+
          R_{\sigma_{i,j+1} \nwarrow \sigma^{\prime}}^{\text{nr}}.
\end{align}
Evidently, in the limit $\lambda \rightarrow 0$ this model reduces to the equilibrium multifarious self-assembly model as defined in~\cite{Murugan_PNAS2015, Sartori_PNAS2020}. 

\section{Results}

\subsection{Shape-shifting structure}

The non-reciprocal multifarious self-organization model lives in a relatively high-dimensional parameter space, 
and therefore, the system can in principle exhibit a plethora of different behaviour. 
Figures \ref{fig3}(a)-(e) show possible outcomes of the model with different input parameters for energy $\varepsilon$, chemical potential $\mu$, and non-reciprocal interaction strength $\lambda$, for $m=4$. A typical successful self-assembly of a desired structure starting from a small initial seed, representing the so-called ``multifarious assembly'' process, is shown in Fig. \ref{fig3}(a). Promiscuous tendency of interacting tiles can potentially result in the growth of a ``chimera'' from an initial seed and hence, leading to unfaithful self-assembly as shown in Fig. \ref{fig3}(b). The ``liquid'' regime is characterized by the presence of tiles in the system without many specific interactions, as shown in Fig. \ref{fig3}(c). The relatively small size of largest ``specifically connected cluster'' (SCC) of the system and the continuous change of the arrangements of the tiles lead to liquid-like structure. When the initial seed is not stable and dissolves, a dilute ``dispersion`` can result, as seen in Fig. \ref{fig3}(d). The instability of the seed can occur either due to a competition between energy and entropy of mixing (as represented by the chemical potential or density) or a strong non-equilibrium activity to due non-reciprocal interactions (see below). Finally, a newly observed ``shape shifter`` regime is shown in Fig.~\ref{fig3}(e), in which appropriate selection of parameters enables the system to switch an initial pattern to the next by exploiting non-reciprocal interactions.

The details of the shape-shifting behaviour is shown with more detail in Fig.~\ref{fig3}(f). Four different structures are stored in the mixture using appropriate choices for the $U^{\text{r}}$ matrix elements, and the corresponding full sequence of length three, namely \{$S^{(1)} \rightarrow S^{(2)}$, $S^{(2)} \rightarrow S^{(3)}$, $S^{(3)} \rightarrow S^{(4)}$\} has been implemented using appropriate choices for the $R^{\text{nr}}$ matrix elements. The simulation starts with introducing a small seed selected from the first structure, which grows and self-assembles to $S^{(1)}$. Subsequently, the non-reciprocal interactions come into effect and cause $S^{(1)}$ to switch to $S^{(2)}$, then proceed to  $S^{(3)}$, and later to $S^{(4)}$. To quantify this feature, Fig. \ref{fig3}(g) shows the overlap of the observed configuration with $S^{(i)}$, defined as $O_i={|A \cap S^{(i)}|}/{|A|}$, where $A=L \cup S^{(i)}$ with $L$ being the largest connected cluster of tiles in the system. This definition takes into account missing tiles or addition of extra tiles. The error is defined as $e_i=1-O_i$, which shows how far the final configuration is from $S^{(i)}$. We observe that for sufficiently large $\lambda$, the system can undergo sequential shifts between the target states.

\subsection{Non-equilibrium drive}

We have systematically explored the parameter space of the non-reciprocal multifarious self-organization model, to uncover the conditions for obtaining the desired behaviour. The  behaviour of the system at equilibrium ($\lambda=0$) is described in Fig.~\ref{fig4}(a), where the error of self-assembly is used to identify the different regimes, namely, chimera, liquid, dispersion, and multifarious self-assembly \cite{Sartori_PNAS2020}. We note that error by itself may not be sufficient to determine the phase of the final configuration, and therefore, it will be important to simultaneously monitor the density, and sometimes, the energy (see SM). 

To explore the effect of non-reciprocal interactions, we plot the error for different values of ($\lambda$-$\mu$) in Fig~\ref{fig4}(b), for fixed $\varepsilon$. 
We observe that for small values of $\lambda$ the self-assembled structures are stable, whereas increasing $\lambda$ beyond a threshold $\lambda_{\rm min}$ introduces the shape shifting behaviour. As expected, the shifts only happen in the multifarious assembly region. When $\lambda$ is larger than a second threshold $\lambda_{\rm max}$, the seed is destabilized and a dispersion is observed. Similar behaviour is observed for fixed $\mu$ 
in the space of ($\lambda$-$\varepsilon$), as shown in Fig.~\ref{fig4}(c).

\subsection{Locating shape shifter region within the multifarious assembly domain}

\begin{figure}
\begin{center}
\includegraphics[width=0.8\columnwidth]{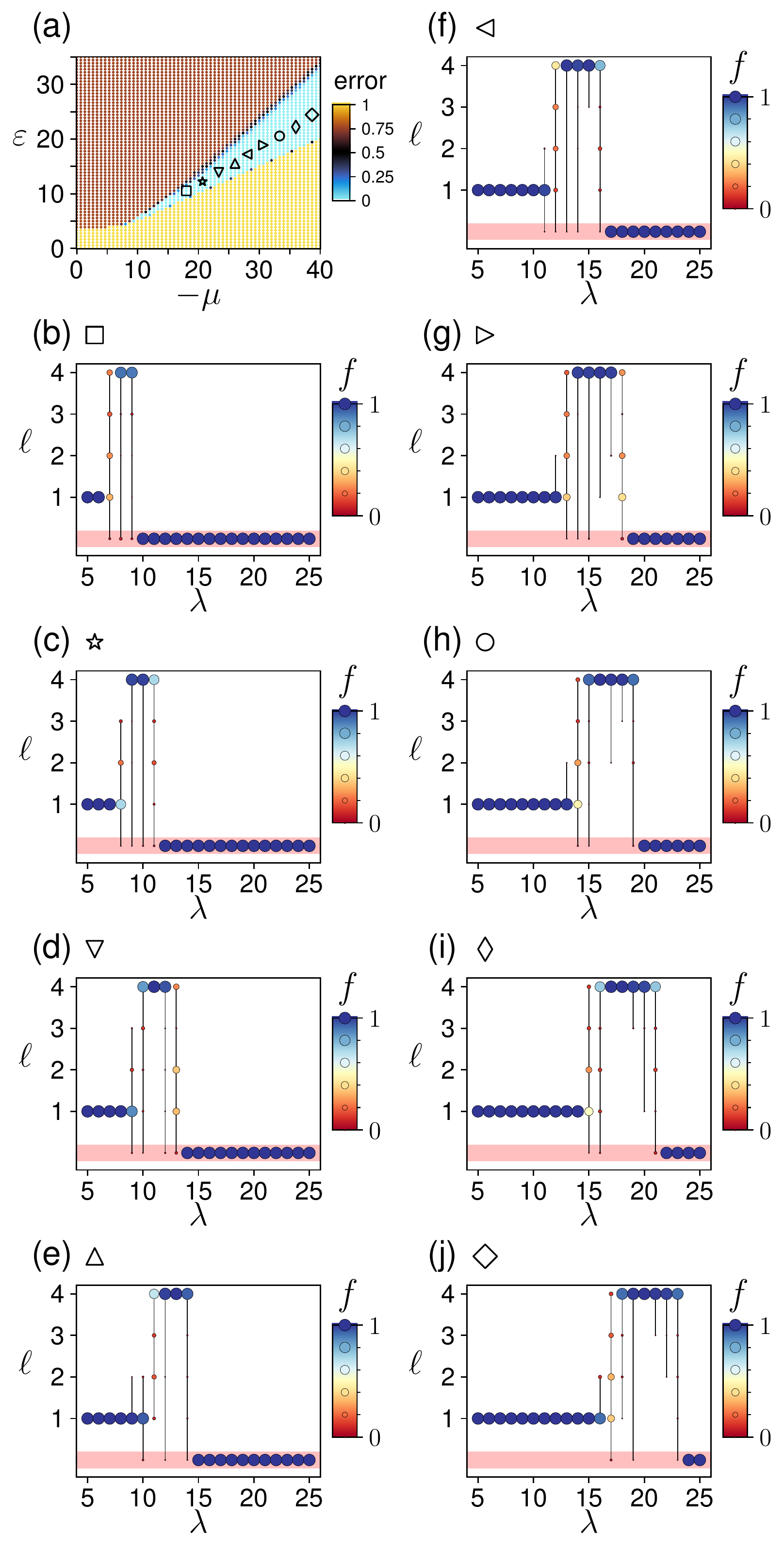}
\caption{Frequency of shifts. The frequency of shifts is characterized for a number of different points in the multifarious assembly regime, as labeled by different markers in (a). The setup corresponds to four desired structures of size~$40\times40$, making a sequence of length 3. Panel~(b) shows the frequency of the shifts for the square marker. Each multi-bulleted bar corresponds to the average of 100 independent realizations. A bullet appears at height $\ell$ if at least one of the realizations ended in the corresponding structure $S^{(\ell)}$. Size and colour of the bullets are related to the frequency or fraction of the realizations that stopped at the corresponding structure at the end of the simulation. The red band corresponds to any erroneous configuration; configurations with $\text{error}>0.1$. Panels~(c)-(j) are the same as (b) but for different ($\varepsilon$-$\mu$) values as marked in (a).}
\label{fig6}
\end{center}
\end{figure}

To shed more light on how the shape shifter domain can be obtained, we present the diagram describing the behaviour of the system in the space of ($\varepsilon$-$\mu$) for different values of $\lambda$, in Fig. \ref{fig5}. Figures~\ref{fig5}(a), (b) correspond to $\lambda=10$, and are based on probing both error and density. As expected, for sufficiently large $\lambda$ the multifarious assembly domain shrinks, as a part of it becomes unstable for the initial seed. 
On the other hand, the remainder of the multifarious assembly domain now accommodates a new shape shifter region, shown in red dashed line in Figs.~\ref{fig5}(a). Increasing $\lambda$ further shrinks continues this trend as shown in Fig.~\ref{fig5}(c).

\subsection{Frequency of shifts}

\begin{figure*}
\begin{center}
\includegraphics[width=0.9\textwidth]{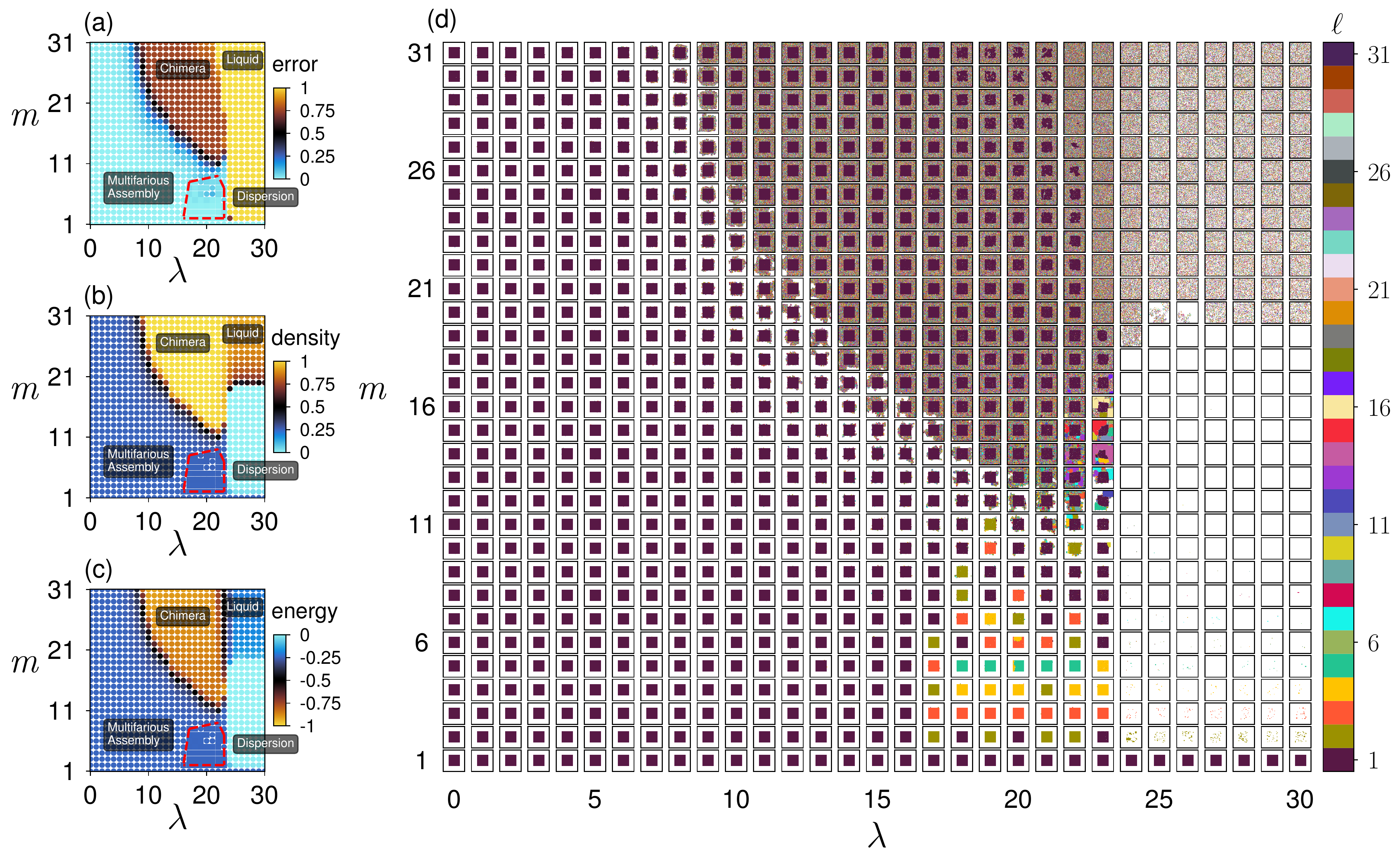}
\caption{ ($m$-$\lambda$) parameter space. The behaviour of the system is probed for different values of desired structure number $m$ and sequence length $m-1$ (with structures of size~$40\times40$). For each set of ($m$,$\lambda$) we have introduced the complete initial structure as seed and simulated the system with 10 independent realizations of the system for $4 \times 10^6 \tau_s$ steps. Panels (a), (b) and (c) show the average error, density, and energy, respectively. Sufficiently strong non-reciprocal interactions introduce the shape shifting behaviour. 
The red dashed line is the convex hull of the points that correspond to $\text{error}<0.05$ and have undergone at least one shift. Panel~(d) shows the last snapshots of the simulation for each ($m$,$\mu$)~pair. 
}
\label{fig7}
\end{center}
\end{figure*}

Let us now characterize the robustness of the shifting behaviour. We focus on the region of interest in the space of ($\varepsilon$-$\mu$), as highlighted in Fig.~\ref{fig6}(a). At each point marked in the multifarious assembly region [Fig.~\ref{fig6}(a)] we change $\lambda\in[5,25]$. To probe the quality of the shifts, we define the frequency $f$ as the fraction of independent realizations of the system that terminated the shifting at one of the configurations along the designed sequence, after a finite simulation time of $4\times10^6 \tau_s$ steps. 
The final configuration can be any of the four states $S^{(1)}$, $S^{(2)}$, $S^{(3)}$, $S^{(4)}$ for any valid shift, or anything else including liquid, chimera, and dispersion. From the functionality point of view {\emph anything else} is an erroneous structure and is not of any use. Panels in Fig.~\ref{fig6} show the frequency at different points in the multifarious assembly region. The red band captures erroneous structures. As observed in the figure, for each marker there is a functional range of $\lambda$ values that can induce the desired shifting between the structures (see SM for more details). 

\subsection{The interplay between capacity and non-equilibrium drive}

\begin{figure}
\includegraphics[width=\columnwidth]{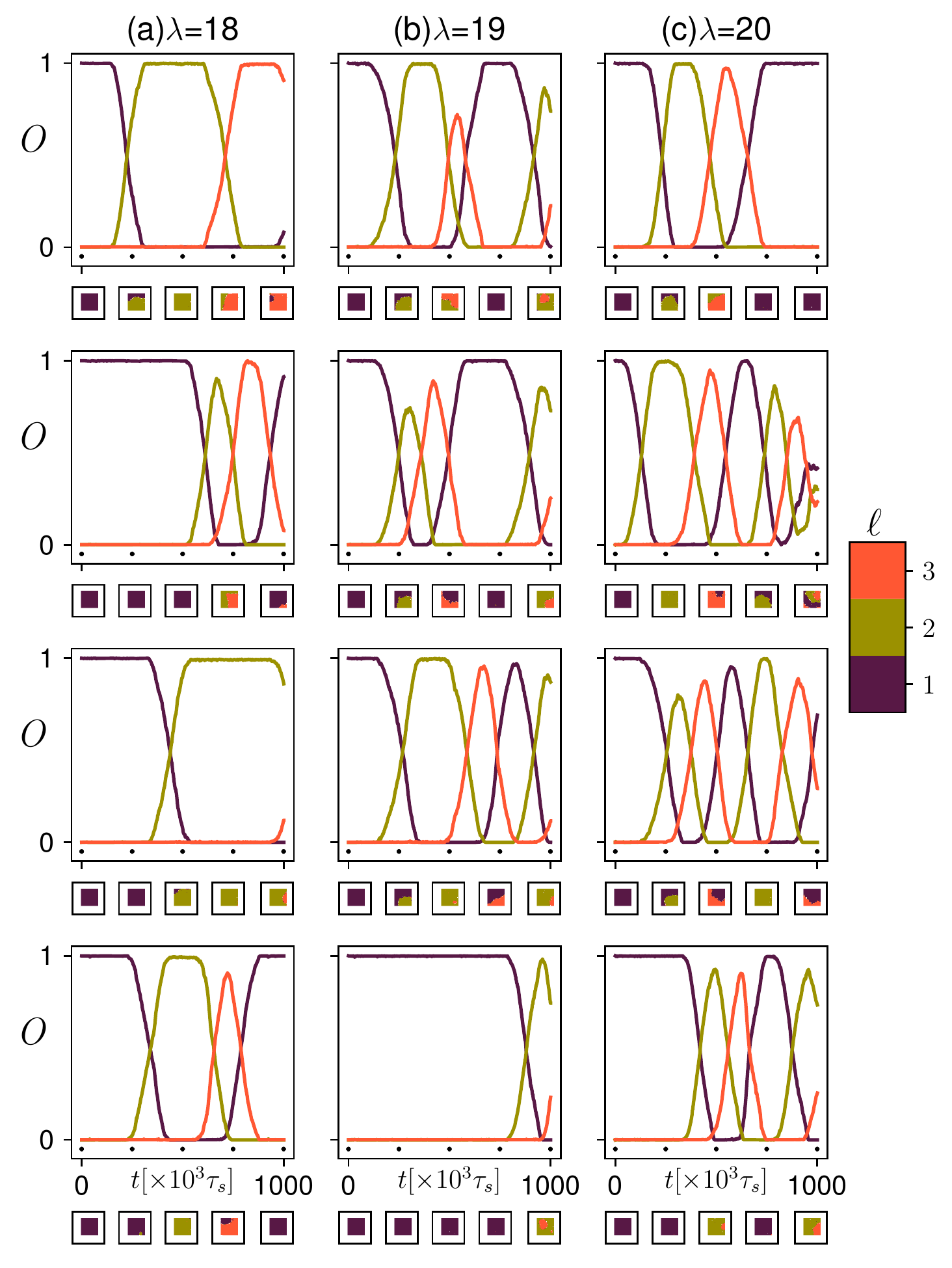}
\caption{Cycles and premature shifting. Here we implement a loop of length three on three predefined structures. For the same set of structures and for three different values of $\lambda$, four different realizations are shown here. ($\varepsilon$-$\mu$) of the system correspond to the $\diamond$ marker in Fig.~\ref{fig6}(a), deep in the multifarious assembly regime. We only plotted the realizations that have undergone at least one shift during the first $10^6 \tau_s$ steps of the simulation.}
\label{fig8}
\end{figure}

It would be interesting to explore what happens when we change the number of stored structures $m$. We have already shown that all the multifarious assembly region potentially provides a design space for shape shifting behaviour appropriate choice of $\lambda$. Let us consider a choice of ($\varepsilon$-$\mu$) point deep in the multifarious assembly regime corresponding to the diamond marker in Fig.~\ref{fig6}(a). For this fixed value of energy and chemical potential, we have mapped out the ($m$-$\lambda$) parameter space, as shown in Fig.~\ref{fig7}. Sequences of different lengths from $m=1$ (a single structure) to $m=31$ (31 structures connected as a sequence of length 30) are stored in the system. The system is then realized independently for different values of $\lambda$, from $\lambda=0$ to $\lambda=30$ from the initial structure, $\ell=1$. For each parameter pair ($m$,$\lambda$) 10 independent realizations of the system is implemented and the three descriptors of error, density and energy, as shown in Figs.~\ref{fig7}(a), (b), and (c), respectively, are used for determining the behaviour of the system. Figure.~\ref{fig7}(d) shows the last of the simulation, which further clarifies the differences between the phases. As we anticipate, weak non-equilibrium drive does not affect the capacity of the system, as it can successfully restore (or keep) the initial seed, and consequently, shape-shifting is not observed. Moreover, we observe the shape-shifting behaviour within the band ($\lambda_{\text{min}}$,$\lambda_{\text{max}}$), and delineated by the red dashed line as convex hull of the points that have undergone at least one shift (see SM for more details).

\subsection{Premature shifting}
Premature shifting refers to the realization of the shifting process from $S^{(\ell)}$ to the next structure $S^{(\ell+1)}$ before the completion of the self-assembly of the current structure $S^{(\ell)}$ in the shifting sequence \{\dots, $S^{(\ell)} \rightarrow S^{(\ell+1)}$, \dots\}. This can happen when in practice the there is no separation of time scale between the process of assembly (driven by the reciprocal interactions) and the shifts between the structures (driven by non-reciprocal interactions). Premature shifting occurs more frequently at larger values of $\lambda$, which reflects the fact that our approximate scheme breaks down at such strongly non-equilibrium conditions.

\subsection{Cyclic shape shifter}

Consider a shifting sequence of length $q-1$ between $q$ structures, \{$S^{(1)} \rightarrow S^{(2)}$, $S^{(2)} \rightarrow S^{(3)}$, \dots, $S^{(q-1)} \rightarrow S^{(q)}$\}. Introducing the shift $S^{(q)} \rightarrow S^{(1)}$ enables the non-reciprocal multifarious self-organization model to run in a loop and realize cyclic structural transitions. Figure~\ref{fig8} shows different realizations of a cyclic shape shifter for three embedded structures and different value of $\lambda$. As $\lambda$ increases, the frequency of shifting between structures increases and besides that premature shifting happens more often. 

\section{Concluding Remarks} 
We have introduced the non-reciprocal multifarious self-organization model, which is capable of both retrieval of stored structures and inducing choreographed transitions between them, hence realizing shape-shifting structures. Through extensive simulations and systematic scanning of the parameter space of the model, we have demonstrated the feasibility of the shape shifter design strategy that is triggered by non-reciprocal interactions. Therefore, we have identified programmable non-reciprocal interactions as a non-equilibrium paradigm using which automated dynamical control of self-assembled structures and transitions between them can be realized.

There have been a number of studies concerning how specific interactions can be designed in colloidal systems such that desired self-assembly routes can be realized experimentally \cite{Zeravcic_RMP2017, Hormoz_PNAS2011, Meng_Science2010}, although these ideas have been explored in the context of protein complexes as well \cite{Sartori_PNAS2020}. On the other hand, catalytically active colloids and enzymes with effective phoretic interactions have been shown to exhibit non-reciprocal interactions \cite{Soto2014,Soto2015,Saha2019,Jaime_PRL2019,OuazanReboul2021}, which are required for the shape-shifting behaviour to emerge. Therefore, using colloidal particles and enzymes as building blocks appears to provide a promising and natural route to the realization of the proposals presented here. 

There are subtleties associated with the practical implementation of the non-reciprocal multifarious self-organization model. One aspect is the reduction of the capacity of the system in comparison to its equilibrium counterpart, which can be remedied by optimizing the trade-off between the heterogeneity of the target structures and sparse usage of the pool, as suggested in Ref. \cite{Sartori_PNAS2020}. The premature shifting between the structures in the queue, mostly for strong non-reciprocal interactions is another challenge, as discussed above. In connection to this issue, we note that the asymmetric Hopfield network model has been shown to have similar limitations, including chaotic behaviour in some circumstances~\cite{Hopfield_PNAS1982, Derrida_1987EPL, Fukai_PRL1990}. In this context, it has been shown that a slow response of the spins can make it possible to circumvent this problem and even lead to the emergence of additional features~\cite{Parisi_1986, Sompolinsky_PRL1986}. This brings the idea that time-dependent specific interactions~\cite{Sahu2009, Zeravcic_PNAS2017} can provide a potential solution to help prevent premature shifts and avoid chaotic behaviour in the self-organization process in our model, and moreover, it may even lead to new phenomena~\cite{Zeravcic_PNAS2017, Zeravcic_RMP2017}. However, this comes with the complexity of the addition of a time-scale to the parameter space of the design. These issues will be addressed in our follow-up studies using appropriate non-equilibrium formalism.

\acknowledgements
We acknowledge support from the Max Planck School Matter to Life and the MaxSynBio Consortium which are jointly funded by the Federal Ministry of Education and Research (BMBF) of Germany and the Max Planck Society.


\bibliography{refs_main}

\end{document}